\documentclass{www-release}

\newcommand{\del}{\textit{del.icio.us}}

\begin{document}

\title{Vocabulary growth in collaborative tagging systems}

\toappear{}

\numberofauthors{2}

\author{
\alignauthor Ciro Cattuto\\
\affaddr{Museo Storico della Fisica e Centro Studi e Ricerche ``Enrico Fermi''}\\
\affaddr{Compendio Viminale, 00184 Roma, Italy}\\
\email{ciro.cattuto@roma1.infn.it}
\alignauthor Andrea Baldassarri\\
\affaddr{Dipartimento di Fisica, Universit\`{a} di Roma ``La Sapienza''}\\
\affaddr{P.le A. Moro, 2, I-00185 Roma, Italy}\\
\email{andrea.baldassarri@roma1.infn.it}
\end{tabular}
\begin{tabular}{c}
\alignauthor Vito D.~P. Servedio\\
\affaddr{Dipartimento di Fisica, Universit\`{a} di Roma ``La Sapienza''}\\
\affaddr{P.le A. Moro, 2, I-00185 Roma, Italy}\\
\email{vito.servedio@roma1.infn.it}	
\alignauthor Vittorio Loreto\\
\affaddr{Dipartimento di Fisica, Universit\`{a} di Roma ``La Sapienza''}\\
\affaddr{P.le A. Moro, 2, I-00185 Roma, Italy}\\
\email{vittorio.loreto@roma1.infn.it}
}

\date{8 February 2007}

\maketitle


\begin{abstract}
We analyze a large-scale snapshot of \del{}
and investigate how the number of different tags in the system grows
as a function of a suitably defined notion of time.
We study the temporal evolution of the global vocabulary size,
i.e.\ the number of distinct tags in the entire system,
as well as the evolution of local vocabularies,
that is the growth of the number of distinct tags used
in the context of a given resource or user. In both cases,
we find power-law behaviors with exponents smaller than one.
Surprisingly, the observed growth behaviors are remarkably regular
throughout the entire history of the system
and across very different resources being bookmarked.
Similar sub-linear laws of growth have been observed in written text,
and this qualitative universality calls for an explanation
and points in the direction of non-trivial cognitive processes in the
complex interaction patterns characterizing collaborative tagging.	
\end{abstract}

\category{H.3.4}{Information Systems}{Systems and Software}
\category{H.3.1}{Information Storage and Retrieval}{Content Analysis and Indexing}


\keywords{folksonomies, collaborative tagging, statistical analysis,\\
growth processes, tag vocabulary, social software, research}


\section{Introduction}
The paradigm of collaborative tagging \cite{mathes2004folksonomies,hammond2005social}
has been swiftly adopted and deployed in a wide range of systems,
motivating a surge of interest in understanding their structure and evolution.
Folksonomies have been known to exhibit striking statistical regularities
and activity patterns \cite{golder2006structure,pnas_tagging}.

In this context, a natural topic for investigation is the
vocabulary of tags that is used within a given system,
and in particular its evolution over time,
as new users, resources and tags come into play.
Some insights in this direction are reported in \cite{golder2006structure}
and \cite{ht06marlow}, but a systematic attempt at characterizing
vocabulary growth in collaborative tagging system is still lacking.
Here we make a first step in that direction by analyzing a large-scale
snapshot of \del{} and identifying a few stylized facts
about the temporal evolution of tag vocabulary in a variety of contexts.

Ordinary vocabularies of words
feature several interesting properties, and one of the most striking
is related to their growth \cite{heapsbook}.
If one scans a text written in natural language and
monitors the number of different words that have appeared as a function of the
total number of words read, one realizes that this growth is described
by a sub-linear law of growth, and often by a power-law behavior with an
exponent smaller than one. It is thus tempting to investigate the
same features in a folksonomy, regarded as a stream of time-ordered posts
in a given context.
How does the number of tags grow? Is the
asymptotic number of tags finite?  What is the rate of invention of new tags?
Does their number eventually reach a plateau?
Beyond the pure theoretical interest, these questions may be important
for collaborative tagging and more generally for understanding
the dynamics of tags in online social communities, where
a deeper understanding of the temporal evolution of the system
is important for both managing existing systems and designing new ones.

The outline of the paper is as follows. Section \ref{sec:data} describes
the experimental data we analyzed. Section \ref{sec:global} is devoted
to the temporal evolution of the global vocabulary, i.e. the growth
of the number of different tags in the entire system, while the analysis
of local vocabulary growth -- the number of distinct tags used
in the context of a given resource or user --
is addressed in Section \ref{sec:local}.
In Section \ref{sec:conclusions} we cast this work in a wider perspective
and discuss some open questions.


\section{Experimental data}
\label{sec:data}
Our analysis will focus on \del{} for several reasons:
i) it was the first system to deploy the ideas and technologies of
collaborative tagging, so it has acquired a paradigmatic character and it is
the natural starting point for any quantitative study. ii) because of its
popularity, it has a large community of active users and comprises a precious
body of raw data on the structure and evolution of a folksonomy.
iii) it is a \textit{broad folksonomy} \cite{vanderwal}, i.e. single tagging events
(posts) retain their identity and can be individually retrieved.
This allows to define and measure the multiplicity (or frequency) of tags
in a given context (for exampe, a resource or a user), providing a precious
opportunity to probe social aspects in the tagging behavior of a community.  
Contrary to this, popular tagging systems falling in the \textit{narrow folksonomy}
class (\textit{Flickr}, for example) are based on a different model of user interaction,
where tags are mostly applied by the content creator, no notion of tag
multiplicity is possible in the context of a single resource,
and no access is given to the raw sequence of tagging events.

The basic unit of information
in a collaborative tagging system is a {\tt(user, resource, \{tags\})} 
triple, here referred to as ``post''.
In \del{} (as well as in many other systems) a post also contains
a timestamp indicating the physical time of the tagging event,
so that the temporal ordering of posts can be preserved
and the dynamical evolution of the system over time
can be reconstructed and investigated.

The dataset used for the present analysis consists of approximately $5 \cdot
10^6$ posts, comprising about $650000$ users, $1.9 \cdot 10^6$ resources and
$2.5 \cdot 10^6$ distinct tags, and covering almost $3$ years of user
activity, from early 2004 up to November 2006. In processing the data,
we discarded all posts containing no tags (about $7$\% of the total).
Regarding tags, since \del{} is case-preserving but not case sensitive, we ignored
capitalization in tag comparison, and counted all different capitalizations of
a given tag as instances of the same lower-case tag.
The timestamp of each post was used to establish post ordering
and determine the temporal evolution of the system.
Posts with invalid timestamps, i.e.\ times set in the future or before \del{} started operating,
were discarded as well (less than $0.5$\% of the total).
After performing the above cuts on the data, a time-ordered sequence of posts
was built and then converted to a time-ordered table of tag assignments (TAS),
by mapping each post of the form {\tt (user, resource, \{tag1, tag2, $\dots$ \})}
into adjacent rows of the form {\tt (tag1, user, resource), (tag2, user, resource), $\dots$ },
one for each tag in the post. Such a table, and selections of it,
were used as the base for analysis described in the following.

Of course, since we rely on post timestamps to reconstruct the history of the
system, our reconstruction is only as much accurate as it is true that posts
are left unchanged after having been entered into the system. We have no way
of detecting and accounting for removed and updated posts, and we will assume
in the following that post removal or updating have a negligible contribution
on the overall evolution of the folksonomy.


\begin{figure}
\includegraphics[width=\columnwidth]{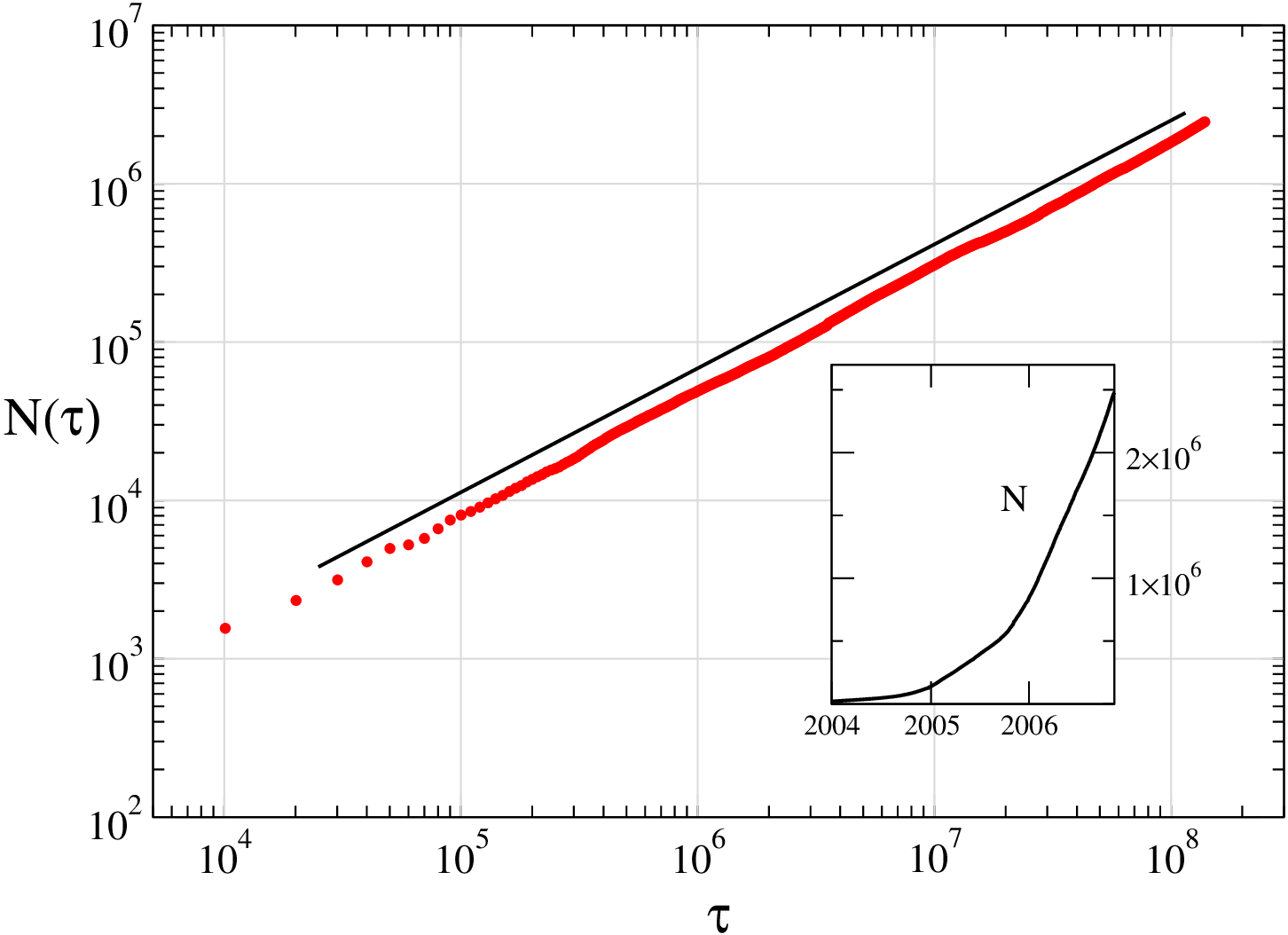}
\caption{Temporal evolution of the total number of distinct tags in \del{}.
As a function of the intrinsic time $\tau$ (see main text), the number $N(\tau)$
of distinct tags (red dots) increases closely following a power-law (straight line
in a log-log plot) across the entire history of the system. The solid black
line, provided as an aid for the eye, corresponds to a power-law with exponent
$\gamma \simeq 0.8$. The inset shows the number $N$ of distinct tags as
function of physical time, spanning almost 3 years of growth
and six orders of magnitude in vocabulary size.
The main graph and the inset refer to the same interval of physical time.
\label{fig:global}}
\end{figure}

\section{Global vocabulary growth}
\label{sec:global}
We begin by studying the evolution over time of the size of the global
``tag vocabulary'', i.e.\ the total number of different tags 
that are present in the folksonomy. As a function of physical time
(inset of Fig.~\ref{fig:global}) the growth of the global vocabulary
is rather featureless, and reflects the huge growth of \del{}
over the past $3$ years. The fact that the system grew up in size
so fast, indeed, makes physical time unsuitable to study the temporal
evolution of \del{}, because a large fraction of the total activity
is compressed in the final part of its temporal history.
Physical time is in many respects ``external'' to the system,
and a much better notion of ``time'' can be defined in terms of quantities
that are intrinsic to the system itself. As mentioned above,
we start our analysis from a time-ordered table of tag assignments.
For the system as a whole, we can define an ``intrinsic time'' $\tau$
as the index of a tag assignment into such a table, so that $\tau$ runs
from $1$ to the number of total tags assignments, i.e. the sum
of the number of tags in all posts (about $1.4 \cdot 10^8$ in our case). 
For each post added to the system, this ``clock'' $\tau$ increases
by a number of ticks equal to the number of tags in that post.

Fig.~\ref{fig:global} shows the total number of distinct tags $N(\tau)$
present in the system at time $\tau$, as a function of $\tau$.
In terms of this intrinsic time, a remarkably clean power-law behavior
(straight line on a log-log plot) can be observed throughout the
full history of the system. This is even more interesting because
the data shown in Fig.~\ref{fig:global} span a time interval covering
almost the entire history of \del: the power-law trend emerges already
at the very beginning and is obeyed all the way to present times,
as the number of active users and that of bookmarked resources
dramatically increase over time. It's worth noticing the following points:
\begin{itemize}
\item
The number $N$ of distinct tags present in the system does not appear
to level off towards a steady-state plateau. This is not surprising
in its own merit because \del{} is an open-ended system and new users
and resources are a source of continuous novelty for the tags
comprised by the folksonomy.
%
\item
The power-law growth followed by $N(\tau)$ is of the form
$N(\tau) \sim \tau^{\gamma}$, with $\gamma < 1$. The black line
in Fig.~\ref{fig:global} corresponds to $\gamma \simeq 0.8$.
%
\item
The rate at which new tags appear at time $\tau$ scales as 
$dN(\tau) / d\tau \sim \tau^{\gamma-1}$. That is, new tags
-- as a function of the intrinsic time $\tau$ -- appear less
and less frequently, with the invention rate of new tags
monotonically decreasing towards zero. The approach to zero is however so slow
that the cumulated number of tags, asymptotically, does not converge
to a constant value but is unbounded ---
assuming the observed trend stays valid.
\end{itemize}
It is remarkable that the above statistical regularities hold
throughout the history of \del{}, while the system undergoes a huge change
in the size of its user base, the number of bookmarked resources,
several changes in the user interface are made, tag suggestion is introduced, and so on.

The above observations constitute the core facts of the present study,
and in the following we will shift from the global view of the system
to a local one, to see whether these facts stay valid, and to deepen our analysis.

\subsection{Sub-linearity in vocabulary growth}
The sub-linear growth reported here is not a newly observed phenomenon.
When dealing with the evolution of the number of attributes pertaining to some
collection of objects, this sub-linear growth is generally referred to as
\textit{Heaps' law} \cite{heapsbook}. As an example, sub-linear behavior has
been observed in the growth of vocabulary size in texts, i.e.\ in the number
of different words in a text as a function of the total number of words
observed while scanning through it. For the case of English corpora,
vocabulary growth exponents in the range $0.4<\gamma<0.6$
have been reported \cite{harman}. The vocabulary
size of the Thai subset of WWW internet web pages has also been found to obey
a sub-linear power-law behavior with exponent $\gamma \approx 0.5$
\cite{sanguanpong}. In contrast, the exponent we observe here is comparatively high.
As a side effect, standard ``approximate text searching'' algorithms might lose efficiency
when applied to folksonomies \cite{baezayates00}. Approximate text searching algorithms
allow a limited number of lexical differences between the terms found and those actually sought.
The key ingredient that allows approximate search algorithms
to scale reasonably is the relatively low vocabulary size,
or equivalently a small exponent $\gamma$. In order to get a feeling of the
scales involved here, consider that in the case of \del{} we have
$\tau \simeq 10^8$, so that the vocabulary size $N(\tau)$ might have been two
order of magnitudes smaller ($10^4$ as compared with $10^6$) if the sub-linear
exponent $\gamma$ had the value $0.5$, which is characteristic of
English texts (instead of the measured $0.8$).
Attempts to explain the power-law behaviors of vocabulary growth
in terms of the measured Zipf's frequency-rank
distribution of words can be found in literature \cite{Leijen05}, as well as
ad-hoc modifications of simple stochastic models \cite{montemurro05}.

It is important to remark that, at odds with texts, no grammatical structure is
embedded in tags. Moreover, the words used in folksonomies are mainly nouns or,
in general, synthetic descriptions of categories~\cite{DBLP:conf/nldb/Veres06}.
In this sense, the only linguistic mechanisms that could be responsible for a
sub-linear growth is a possibly hierarchical organization of tags
induced by semantics. Another important difference is that in written texts
the number of authors is usually limited, while the number of
users contributing to the tag vocabulary of \del{} is large and growing in time.


\newpage

\section{Local vocabulary growth}
\label{sec:local}
We will now shift our focus to a local scope,
moving from a global view of tag vocabulary to a more fine-grained one,
dealing with the restricted contexts of single resources and users.
Specifically, we will investigate how the number of different tags associated
with a given resource (or user) grows as a function of an intrinsic time.
The notion of time we adopt in the following is the same
we employed for the global analysis of Section \ref{sec:global},
except that in this case it is restricted to the context of a single
resource or user: given a resource (or a user),
we select from the global, time-ordered TAS table
only those tag assignments that involve that resource (or user).
We define the intrinsic time $\tau$ as the index into the selection,
i.e. the cumulated number of tags associated with that resource (or user).
Thus our notion of time is resource-dependent (or user-dependent),
and $\tau$ naturally measures metadata accumulation in the specific
semantic context of a single resource or user.

\begin{figure}
\includegraphics[width=\columnwidth]{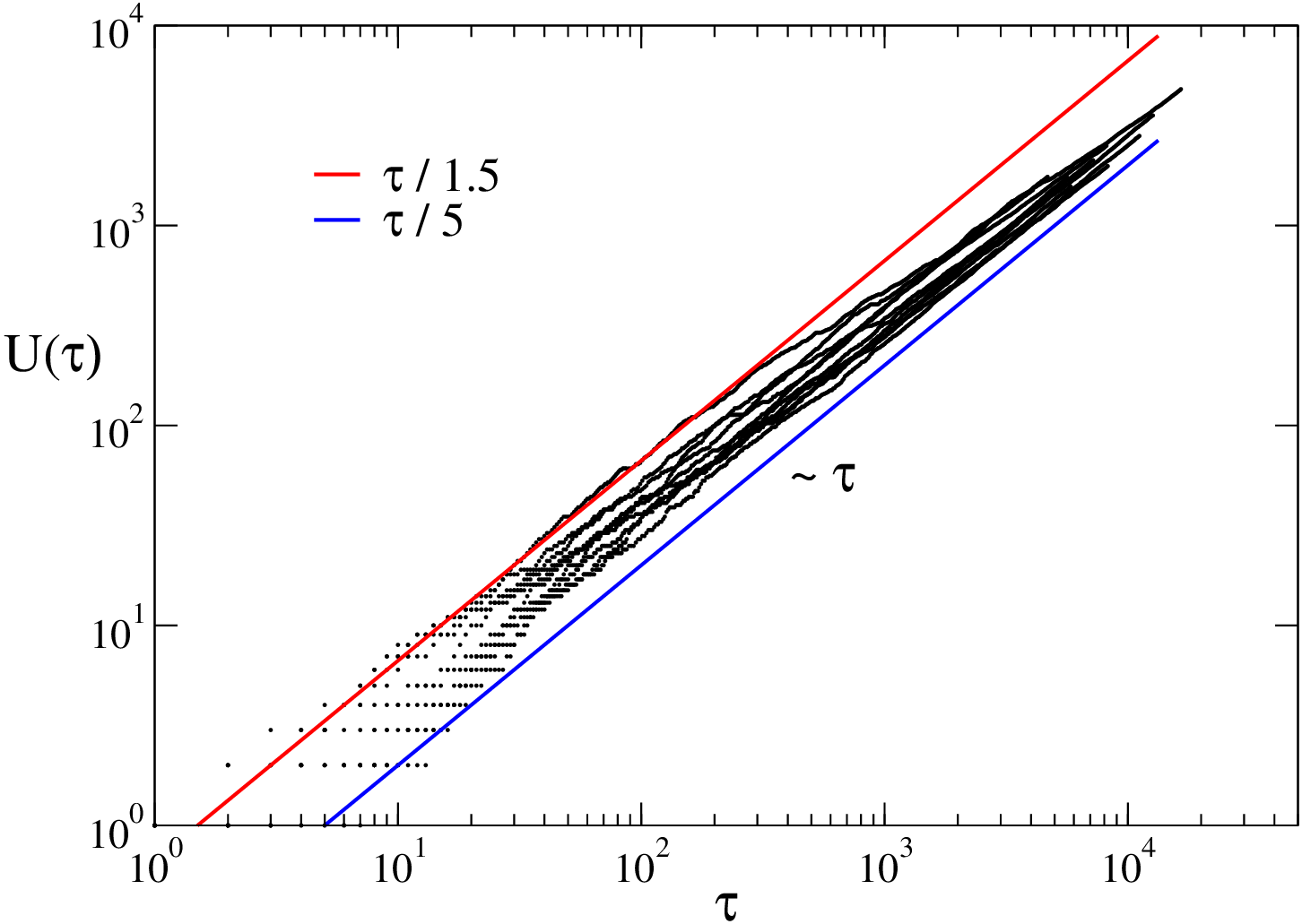}
\caption{For $10$ different resources in \del{}, the number of users $U(\tau)$
who have bookmarked them is shown as a function of the intrinsic time $\tau$
pertaining to each resource. The $10$ chosen resources are
among the $1000$ top-bookmarked resources in \del{}, with ranks starting from
$100$ and decreasing at intervals of $100$.
An approximately linear trend is visible, which correspond to a comparatively stable average number of tags
in the posts associated with those resources: the blue and red lines
show the expected linear growth for two different number of
average tags per post, respectively 1.5 and 5.
\label{fig:user10}}
\end{figure}

\begin{figure}
\center
\includegraphics[width=0.8\columnwidth]{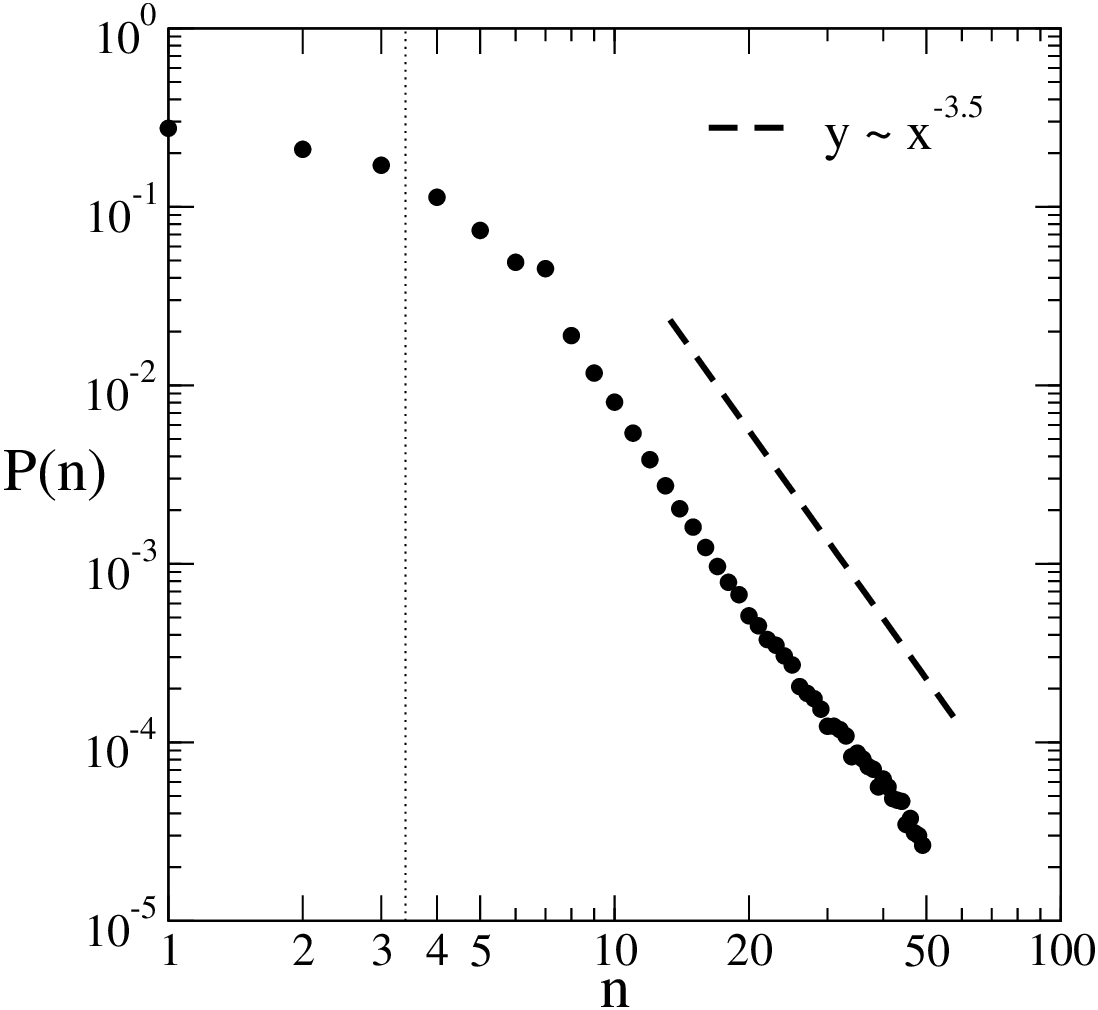}
\caption{Probability distribution $P(n)$ of the number of tags
contained in a post in \del{}. The distribution displays a steep
power-law tail ($\sim n^{-3.5}$, dashed line), and the global average value
$\bar{n}$ is well-defined, with $\bar{n} \simeq 3.4$ (vertical line).
\label{fig:post_len}}
\end{figure}

\subsection{Resource-specific vocabularies}
\label{sec:local:res}
In Fig.~\ref{fig:user10} we consider $10$ different popular resources
and we plot the number of users who have bookmarked them as a function of the
intrinsic time $\tau$ for each resource (the total number of tags assigned to
it). The resources are chosen among the $1000$ top-bookmarked resources in the
system, starting from rank $100$ and decreasing at intervals of $100$.
The growth behaviors are approximately lineare and quite homogeneous,
and no systematic differences are observed with respect to the rank, at this level.
The observed linear dependence can be easily understood by studying
the probability distribution of the number of tags contained in a post,
here referred to as ``post length''.
The global distribution of the number of tags in a post
is shown in Fig.~\ref{fig:post_len}.
It is interesting to notice that the distribution displays
an initial exponential decay, with a typical number of tags equal to $3$-$4$,
while for large post lengths it becomes algebraic.
The power-law tail, however, is quite steep,
with an exponent that appears to be close to $-3.5$.
This means that such a fat tail does not contribute significantly
to the dispersion of the post length when the number of posts is increased,
i.e. that the average number of tags $\bar{n} \simeq 3.4$ is an intrinsic
characteristic of the system and does not depend strongly on the size of the dataset.
Because of this, in the context of a resource, each post contributes
exactly $1$ new user and a well-defined number of tags $\bar{n}$
(eventually adding up to $\tau$), so that in Fig.~\ref{fig:user10} a linear dependence
is observed between the cumulated number of users $U(\tau)$
and $\tau$. \footnote{It is interesting to notice that posts with more
than $40$ tags are present, and they are not just due to spammers.}

\begin{figure}
\includegraphics[width=\columnwidth]{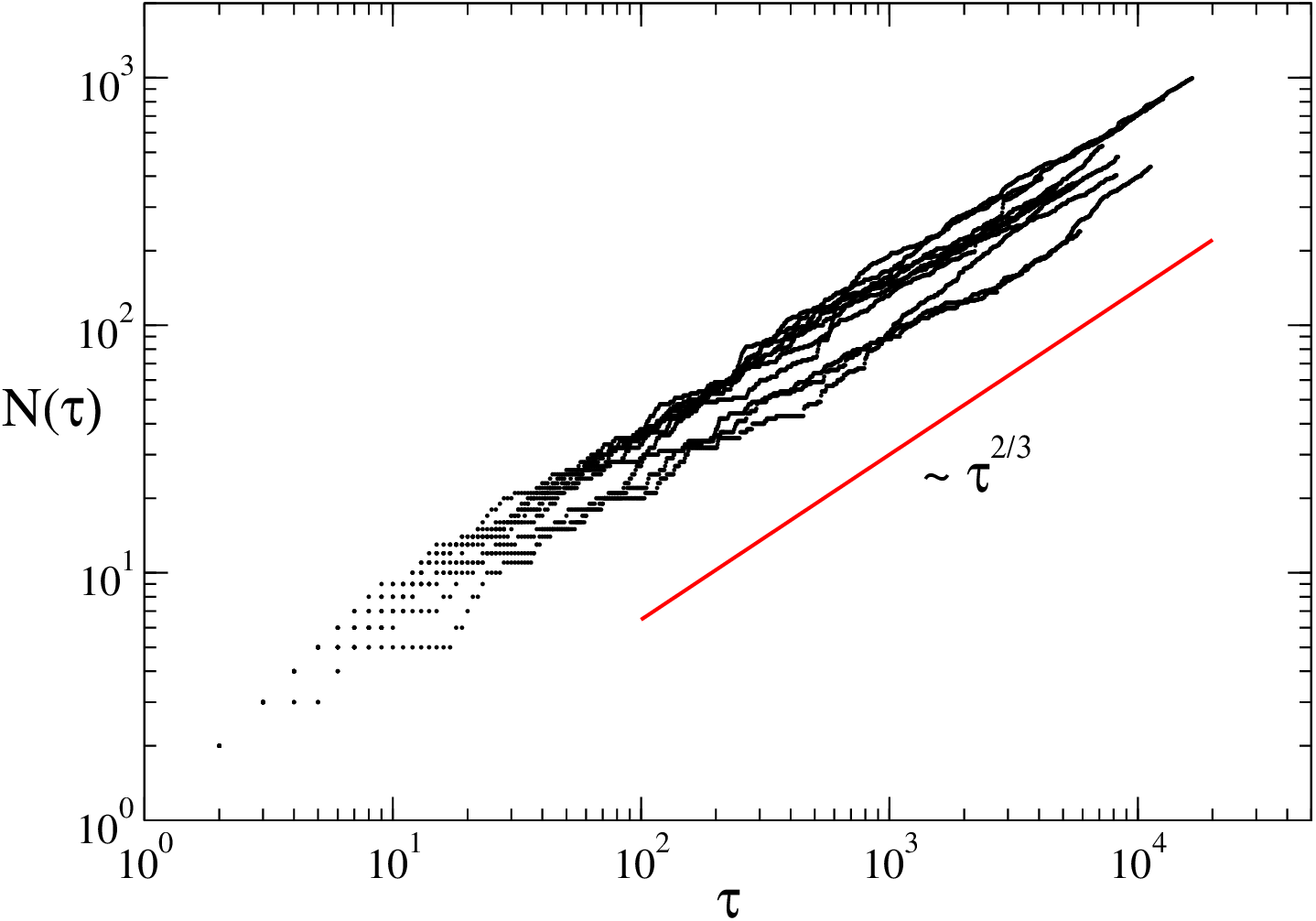}
\caption{Vocabulary growth for single resources.
For $10$ different resources in \del{}, the number of distinct tags $N(\tau)$
associated with them is plotted as a function of the
intrinsic time $\tau$ pertaining to each resource. 
The $10$ chosen resources are the same of Fig.~\ref{fig:user10}.
While single resources display a
somehow noisy evolution, an overall power-law behavior governing the
vocabulary growth is apparent, with an exponent $\gamma \simeq 2/3$ (red line).
\label{fig:res10}}
\end{figure}

\subsection{Scaling}
\label{sec:local:scaling}
The large number of users tagging each resource make the statistical
features of resources quite similar, as long as they are popular enough.
This can be shown, for instance,
by looking at the vocabulary growth in the context of a single resource.
To this end, we consider the growth of the number
$N(\tau$) of distinct tags associated with the same $10$ popular resources of
Fig.~\ref{fig:user10}, as a function of the intrinsic (resource-specific) time
$\tau$. While the vocabulary growth exhibits a somehow noisy temporal
evolution, the general trend of growth appears to be compatible with an algebraic
law of growth, a power-law with an exponent close to $2/3$.
This is a striking regularity, valid for very different resources across the system.
Also, at this level of detail, no systematic dependence on the popularity of a
resource can be detected. The local exponent of growth is smaller than the
global one (Fig.~\ref{fig:global}) and the relation between the two
may be linked to the statistical properties of tag co-occurrence, and might
ultimately provide insights into the semantic structure of folksonomies.

To better probe the similarity of growth behaviors for different resources,
we defined a rescaled growth curve, where both the intrinsic time $\tau$
and the final number of distinct tags $N(\tau_{max})$ are divided
by their final values, $\tau_{max}$ and $N(\tau_{max})$, respectively.
In this way, the curves for different resource can be easily plotted on the same graph.
As shown in Fig.~\ref{fig:collapse}, all the rescaled curves lie between
two limit power-laws, $(\tau/\tau_{max})^{1}$ and $(\tau/\tau_{max})^{1/2}$.
More importantly, all curves tend to lie along a ``universal'' growth curve
with an exponent close to $2/3$.

\begin{figure}
\includegraphics[width=\columnwidth]{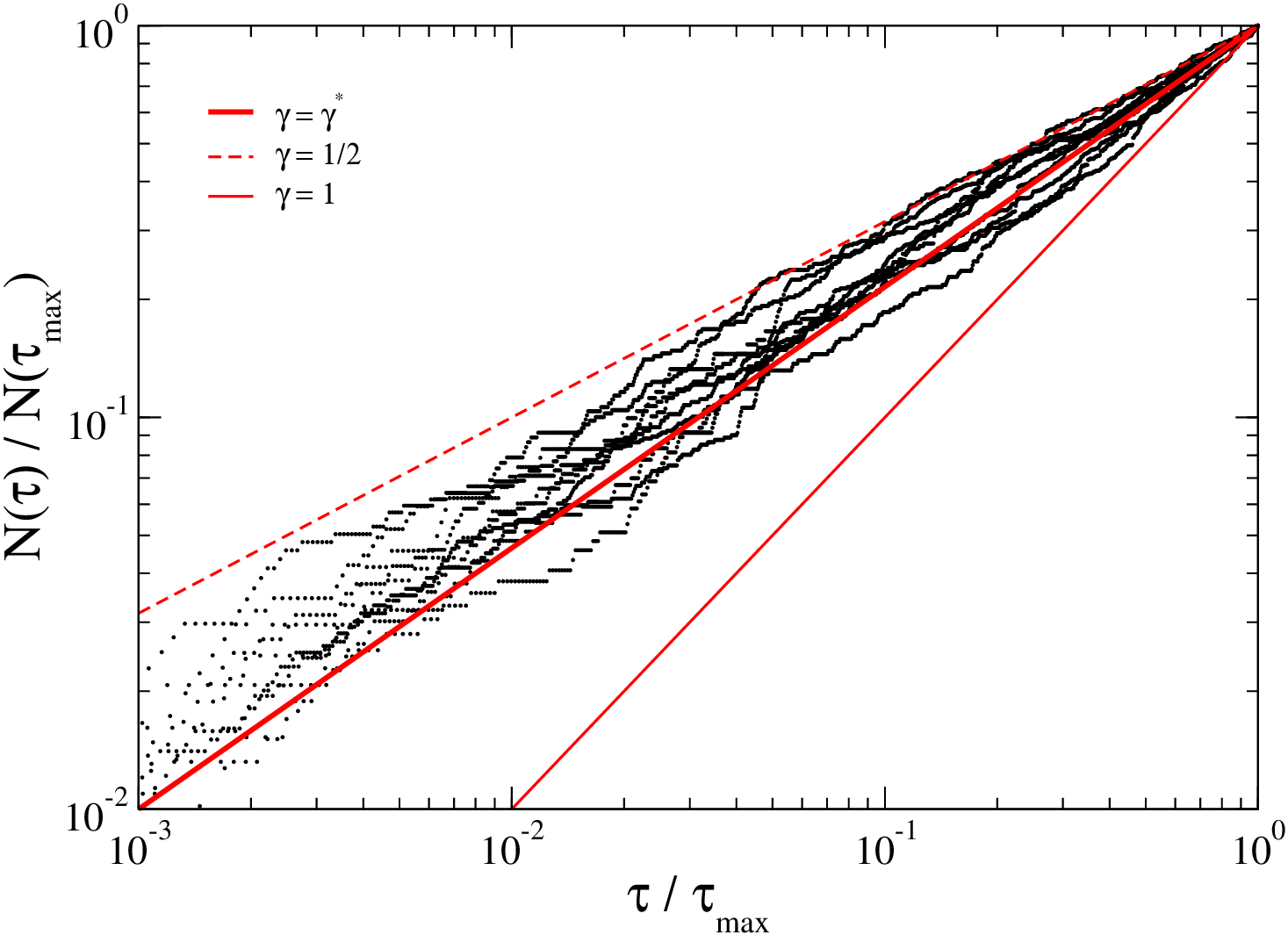}
\caption{Rescaled vocabulary growth. The curves of Fig.~\ref{fig:res10}
were rescaled by dividing both the intrinsic time $\tau$
and the number of distinct tags $N(\tau)$ by their final (resource-specific)
values $\tau_{\mbox{\small max}}$ and $N(\tau_{\mbox{\small max}}$, respectively.
After rescaling, all curves lie approximately
along the ``universal'' $(\tau / \tau_{\mbox{\small max}})^{2/3}$ line
(thick red line). On approaching the common endpoint,
the slope of all curves appear to lie in the $0.5$-$1$ range
(dashed line and thin red line, see also Fig.~\ref{fig:exp}).
\label{fig:collapse}}
\end{figure}

\subsection{Distribution of growth exponents}
\label{sec:local:exp}
In order to make a more quantitative measure over a broader set of resources,
we implement the following unsupervised procedure for characterizing
the growth of local tag vocabularies: for each resource we measure
an effective exponent $\gamma$ that approximates
the rescaled vocabulary growth with a power-law $(\tau/\tau_{max})^\gamma$.
The simplest way to do this is to
compute $\gamma$ as $\gamma=\log(N(\tau_{max}))/\log(\tau_{max})$.
Fig.~\ref{fig:exp} shows the probability distribution of the resulting
values of $\gamma$, measured for three groups of resources.
In particular, the red curve in Fig.~\ref{fig:exp} displays
the distribution of $\gamma$ values for the $1000$ top ranked
(most bookmarked) resources in \del. The distribution is well
approximated by a rather narrow Gaussian distribution,
whose average value is $\gamma^{*} \simeq 0.7$.
This seems to confirm the idea (Fig.~\ref{fig:collapse}) that there is
a well-defined exponent of growth governing the temporal evolution
of popular resources. Moreover, the vocabulary growth of popular resources
appears slower than the system-wide vocabulary growth of Fig.~\ref{fig:global}.

\begin{figure}[t]
\includegraphics[width=\columnwidth]{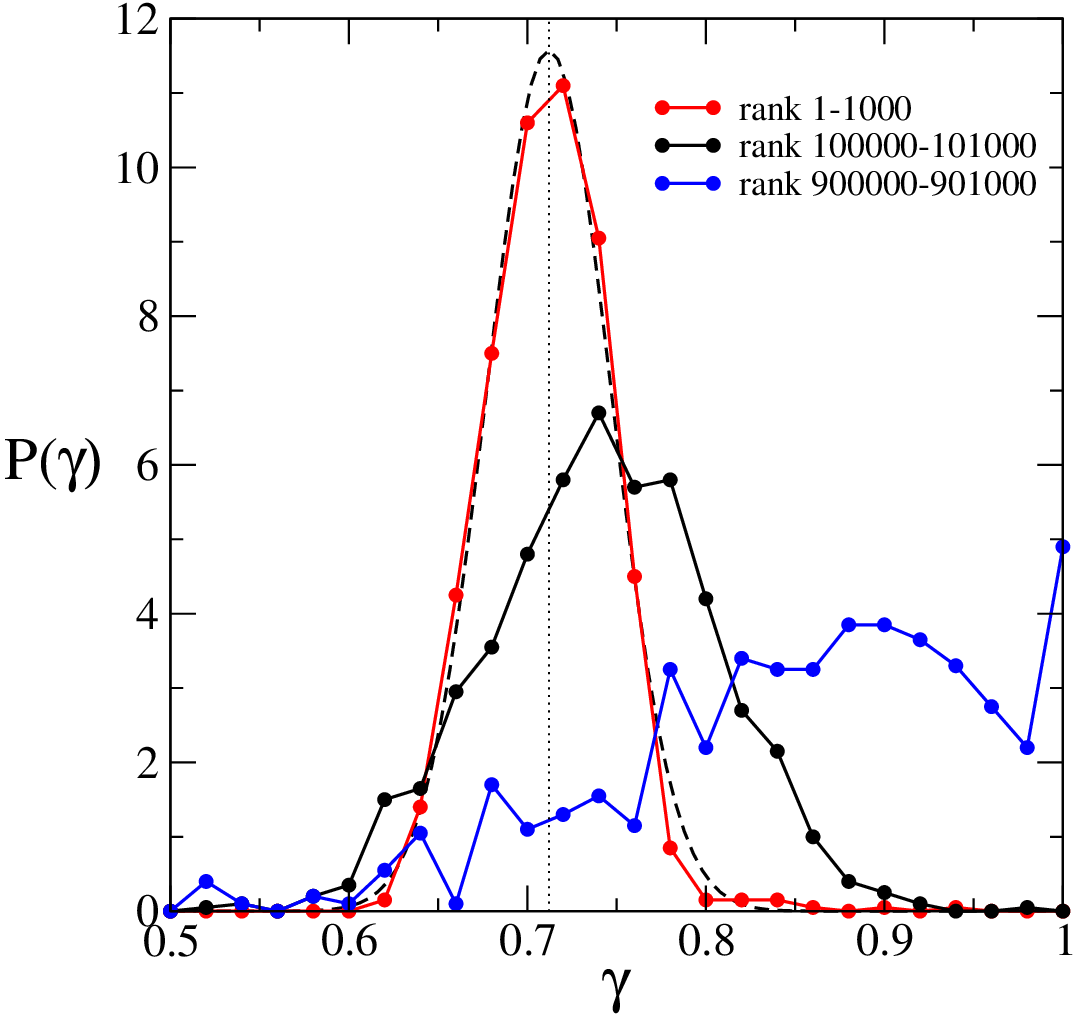}
\caption{Probability distribution of the vocabulary growth exponent $\gamma$
for resources, as a function of their rank. The red curve is the normalized
probability distribution $P(\gamma)$ for the $1000$ top-ranked
(most bookmarked) resources in \del{}. It appears to be sharply peaked
at a characteristic value $\gamma^{*} \simeq 0.71$ (vertical line)
and can be closely fitted with a Gaussian (dashed line).
This indicates that highly bookmarked
resources share a characteristic law of growth, as already pointed out
in Fig.~\ref{fig:collapse}. On computing the distribution $P(\gamma)$
for less and less popular resources (black curve and blue curve),
the peak shifts towards higher values of $\gamma$
and the growth behavior becomes more and more linear.
The typical number of users who have bookmarked the resources used
in this analysis is approximately a few thousands for the red curve,
a few hundreds for the black curve, and just a few users for the blue one.
\label{fig:exp}}
\end{figure}

On computing the distribution $P(\gamma)$
for less and less popular resources (black and blue curves),
the distribution gets broader and its peak shifts towards
higher values of $\gamma$, indicating that the growth
behavior is becoming more and more linear. This crossover
from sub-linear to linear growth for resources bookmarked
by just a few users is expected and corresponds
to a sort of ``priming'' effect for the resource:
the first few users who bookmark it build the ``core''
tag vocabulary for the resource, and since only a few posts
are present at that time, most tags are new and the
size of the vocabulary grows linearly with the total
number of tags $\tau$ as well as with the number of posts
associated with the resource. As more and more users
bookmark the resource, correlations and social effects
come into play and the law of growth crosses over
from the linear to the ``universal'' sub-linear behavior reported above.

To make contact between local vocabulary growth in the context
of a single resource and vocabulary growth in the context of a single user,
we repeat the above analysis for the $1000$ most active users
in \del{} (as measured by the number of resources they bookmarked).
The resulting probability distribution $P(\gamma)$ is shown
in Fig.~\ref{fig:users} and is qualitatively similar
to the ones of Fig.~\ref{fig:exp}. In particular,
we notice that the peak of $P(\gamma)$ is compatible
with the value $\gamma^{*}$ observed for the top-ranked resources.

We would like to remark that the huge variability
of vocabularies, at the level of single users and resources,
is not in contrast with very regular -- and simple -- features
at the global level. On the contrary, the emergence of regularity
from the uncoordinated activity of users is the hallmark of complexity
and indicates that tools and concepts from complex system science
may prove valuable for understanding the structure and dynamics of folksonomies.

\begin{figure}
\includegraphics[width=\columnwidth]{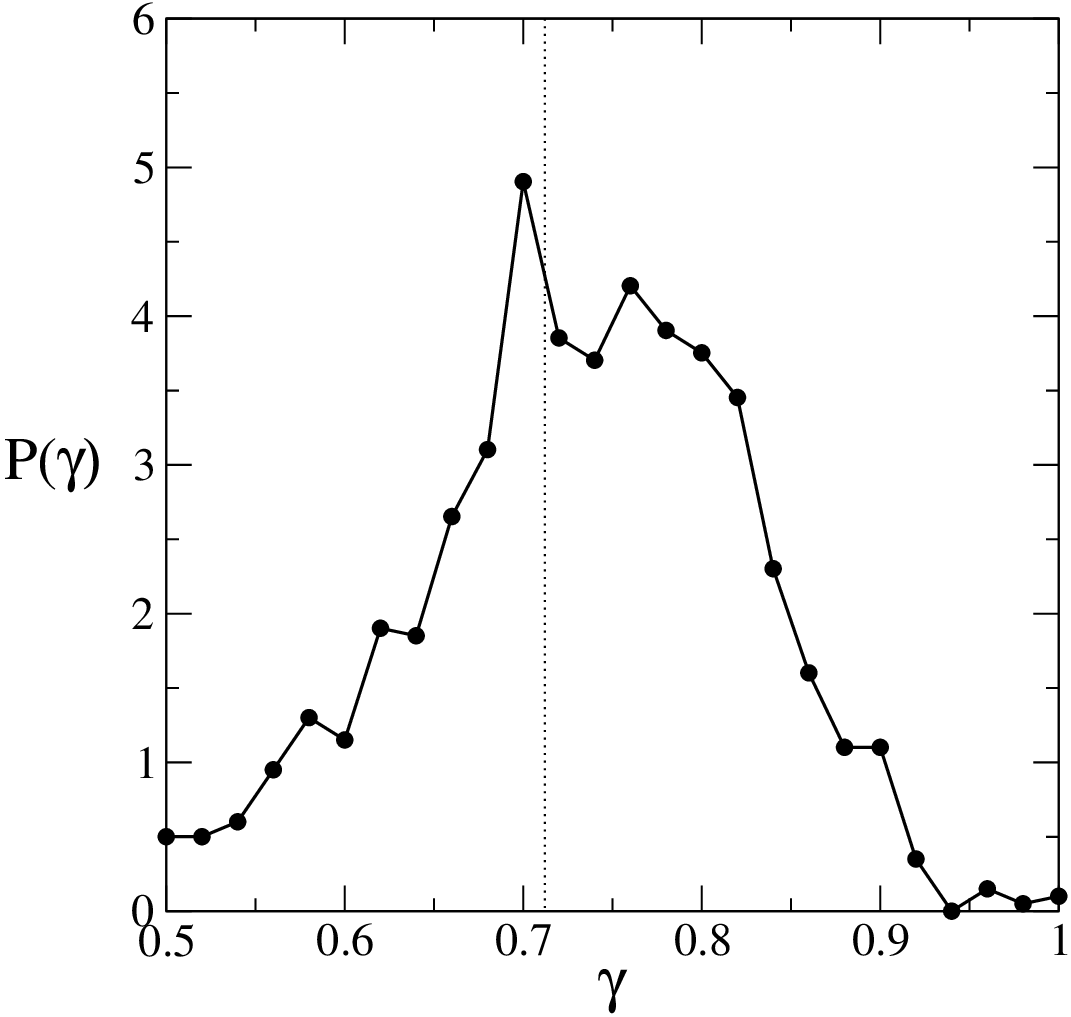}
\caption{Probability distribution of the vocabulary growth exponent $\gamma$
for user vocabularies. The distribution $P(\gamma)$ was computed
for the $1000$ most active users in \del{}. Similarly to Fig.~\ref{fig:exp},
it appears peaked around a characteristic value close to the same
observed for top-ranked resources (vertical line, same as in Fig.~\ref{fig:exp}). 
\label{fig:users}}
\end{figure}


%
%

\section{Conclusions}
\label{sec:conclusions}
In this paper we have presented a statistical analysis of a large-scale
snapshot of \del. We focused in particular on the growth of the system
as mirrored by the number of distinct tags present in the system at a given time.
We introduced a notion of intrinsic time, based on metrics that are internal
to the system itself. In contrast with physical time, this definition exposes
the natural laws of growth of the system and overcomes the trivial bias
due to the growth of the user base. We investigated the growth of the
global tag vocabulary as well as the growth of local vocabularies
in the context of a given resource or user.

A first interesting result is a power-law growth -- characterized by an exponent smaller
than one -- of the number of distinct tags at the global level.
This growth displays the same functional form, with no discontinuities, throughout
the entire history of \del. This is a surprising result,
especially when considering the open-ended nature of the system
and the several changes that have occurred in the interaction between the users
and the system.


Analyzing the growth dynamics of local vocabularies,
associated with a given resource or user, may provide insights
into the relationship between the behavior of individual users and
vocabulary growth at the system -- or community -- level,
as well as insights into the process of invention of new tags.
For popular resources in \del{} we report a sub-linear growth
with exponents sharply peaked around a characteristic value
(slightly different from the global one), while for less popular resources
we observe exponent values slowly shifting towards $1$.
The sub-linear growth observed at the local level cannot be explained
as a mere reflection of the growth in the number of users,
(which, for resources, is linear in the intrinsic time)
nor as an increase in the average number of tags per post
(which has a rather stable characteristic value).

These observations point out that sub-linear dictionary growth is a genuine
non-trivial feature of the system and open several problems.
Is sub-linear growth at the global level
(or at the local level) related to correlations among users' activity?
Does the growth observed in the context of a single user
reflect a collective/cooperative phenomenon,
or is it just mirroring the complex cognitive processes (incorporating semantics)
at the level of that individual user?
Is the difference between local and global exponents relevant,
and if so, what kind of information about the structure of tag space is it disclosing?
What are the key elements in the user-system interaction that lead to the observed behaviors?

These questions may play an important role for applications as well,
especially in terms of defining quality metrics for the emergent vocabulary of a folksonomy,
both at the global level and in semantically narrower contexts.
Since the breadth of the tag vocabulary is linked to navigability,
this might eventually impact the design of new systems.


\newpage

\section{Acknowledgments}
This research has been partly supported by the
TAGora project (\texttt{FP6-IST5-34721}) funded by the Future and Emerging                   
Technologies program (IST-FET) of the European Commission. 
The information provided is the              
sole responsibility of the authors and does not reflect the                     
Commission's opinion. The Commission is not responsible for any use             
that may be made of data appearing in this publication.


%
%


\end{document}